\documentclass[aps,prl,reprint,groupedaddress]{revtex4-1}

\usepackage{graphicx}
\usepackage{dcolumn}
\usepackage{bm}
\usepackage{hyperref}

\newcommand{\ah}{\hat a}
\newcommand{\cO}{{\cal O}}
\newcommand{\wh}{\widehat}
\newcommand{\nn}{\nonumber}
\newcommand{\eqn}[1]{(\ref{#1})}
\newcommand{\svs}{\vbox{\vskip 4mm}}

\newcommand{\MSb}{{\overline{\rm MS}}}
\newcommand{\sfrac}[2]{\mbox{$\frac{#1}{#2}$}}

\begin{document}


\title{Scheme variations of the QCD coupling and  hadronic
 \boldmath{$\tau$} decays}


\author{Diogo Boito}
\affiliation{Instituto de F\'isica de S\~ao Carlos, Universidade de
             S\~ao Paulo, CP 369, 13560-970, S\~ao Carlos, SP, Brazil}

\author{Matthias Jamin}
\affiliation{IFAE, BIST, Campus UAB, 08193 Bellaterra (Barcelona) Spain}
\affiliation{ICREA, Pg.~Llu\'\i s Companys 23, 08010 Barcelona, Spain}

\author{Ramon Miravitllas}
\affiliation{IFAE, BIST, Campus UAB, 08193 Bellaterra (Barcelona) Spain}


\begin{abstract}
The Quantum Chromodynamics (QCD) coupling, $\alpha_s$, is not a physical
observable of the theory since it depends on conventions related to the
renormalization procedure. We introduce a definition of the QCD coupling,
denoted by $\wh\alpha_s$, whose running is explicitly renormalization scheme
invariant. The scheme dependence of the new coupling $\wh\alpha_s$ is
parameterized by a single parameter $C$, related to transformations of the QCD
scale $\Lambda$. It is demonstrated that appropriate choices of $C$ can lead
to substantial improvements in the perturbative prediction of physical
observables. As phenomenological applications, we study $e^+e^-$ scattering
and decays of the $\tau$ lepton into hadrons, both being governed by the QCD
Adler function.
\end{abstract}


\maketitle


Perturbation theory in the strong coupling, $\alpha_s$, is one of the
central approaches to predictions in Quantum Chromodynamics (QCD). Because of
confinement, however, $\alpha_s$ is not a physical observable: its definition
inherently depends on theoretical conventions such as renormalization scale and
renormalization scheme. Obviously, measurable quantities should not depend on
such choices. Regarding the renormalization scale, this independence condition
allows to derive so-called renormalization group equations (RGE) which have to
be satisfied by all physical quantities. For the renormalization scheme, the
situation is more complicated, because order by order the strong coupling can
be redefined. For that reason, perturbative computations are performed mainly
in convenient schemes like minimal subtraction (MS)~\cite{hv72} or modified
minimal subtraction ($\MSb$)~\cite{bbdm78}.

The aim of this work is to introduce a new definition of the strong coupling,
$\wh\alpha_s$, that satisfies two properties. First, the scale running of the
coupling, described by the $\beta$-function, is explicitly scheme invariant.
Second, the scheme dependence of the coupling can be parameterized by a single
parameter $C$. Hence, in the following, we shall refer to this scheme as the
$C$-scheme, even though we are actually considering a whole class of schemes.
Variations of $C$ will directly correspond to transformations of the QCD scale
parameter $\Lambda$.

We then proceed to apply our coupling definition to concrete cases. Among
the best studied QCD quantities to which the $C$-scheme may be applied is the
two-point vector correlator and the related Adler function~\cite{adl74}, which
emerge in calculations of the total cross section of $e^+e^-$ scattering into
hadrons, and that also govern theoretical predictions of the inclusive decay
rate of $\tau$ leptons into hadronic final states~\cite{bnp92}. At present,
their perturbative expansion is known up to the fourth order in
$\alpha_s$~\cite{bck08}. Having at our disposal a parameter to investigate
scheme variations, we show that appropriate choices of $C$ can lead to
substantial improvements in the predictions for these quantities. The use of
$\wh\alpha_s$ for the scalar correlator, which is relevant for the prediction
of Higgs boson decay into quarks and for light quark-mass determinations from
QCD sum rules,  is investigated in a related article~\cite{jm16}.

Compared to other celebrated methods used for the optimization of perturbative
predictions, the procedure we present here differs in more than one way. The
main difference is that we seek to optimize the perturbative prediction by
exploiting its scheme dependence, while the idea behind methods such as
BLM~\cite{Brodsky:1982gc} or PMC~\cite{Brodsky:2012rj,Mojaza:2012mf} is to
obtain a scheme-independent result through a well defined algorithm for setting
the renormalization scale, regardless of the intermediate scheme used for the
perturbative calculation (which most often is $\MSb$). Furthermore, some of
these methods, such as for example the
``effective charge"~\cite{Grunberg:1982fw}, involve a process dependent
definition of the coupling. In the procedure described here, one defines a
process independent class of schemes, parameterized by a single continuous
parameter $C$.  We then explore variations of this parameter in order to
optimize the perturbative series in the spirit of asymptotic expansions. This,
however, entails that preferred values of the parameter $C$ depend on the
process considered.

\vspace{2mm}

Let us begin with the scale running of the QCD coupling $\alpha_s$, which is
described by the $\beta$-function as
\begin{equation}
\label{bfun}
-\,Q\,\frac{{\rm d}a_Q}{{\rm d}Q} \,\equiv\, \beta(a_Q) \,=\,
\beta_1\,a_Q^2 + \beta_2\,a_Q^3 + \beta_3\,a_Q^4 + \ldots \,. 
\end{equation}
Here and in the following, $a_Q\equiv\alpha_s(Q)/\pi$, $Q$ is a physically
relevant scale, and the first five $\beta$-coefficients $\beta_1$ to $\beta_5$
are analytically available \cite{lrv97,bck16}. (In our conventions, they have
been collected in Appendix~A of Ref.~\cite{jm16}.)

Employing the RGE \eqn{bfun} for $a_Q$, the well-known scale-invariant QCD
parameter $\Lambda$ can be defined by
\begin{equation}
\label{Lambda}
\Lambda \,\equiv\, Q\, {\rm e}^{-\frac{1}{\beta_1 a_Q}}
\,[ a_Q ]^{-\frac{\beta_2}{\beta_1^2}}
\exp\Biggl\{\,\int\limits_0^{a_Q}\,\frac{{\rm d}a}{\tilde\beta(a)}\Biggr\}\,,
\end{equation}
where
\begin{equation}
\frac{1}{\tilde\beta(a)} \,\equiv\, \frac{1}{\beta(a)} - \frac{1}{\beta_1 a^2}
+ \frac{\beta_2}{\beta_1^2 a} \,,
\end{equation}
which is free of singularities in the limit $a\to 0$. Let us consider a scheme
transformation to a new coupling $a'$, which takes the general form
\begin{equation}
\label{ap}
a' \,\equiv\, a + c_1\,a^2 + c_2\,a^3 + c_3\,a^4 + \ldots \,.
\end{equation}
The $\Lambda$-parameter in the new scheme, $\Lambda'$, only depends on $c_1$
and not on the remaining higher-order coefficients. The precise relation reads~\cite{cg79}
\begin{equation}
\label{Lambdap}
\Lambda' \,=\, \Lambda\,{\rm e}^{c_1/\beta_1} \,.
\end{equation}

The fact that redefinitions of the $\Lambda$-parameter only involve a single
constant motivates the implicit definition of a new coupling $\ah_Q$, which is
scheme invariant, except for shifts in $\Lambda$, parameterized by a parameter
$C$:
\begin{eqnarray}
\label{ahat}
\frac{1}{\hat a_Q} + \frac{\beta_2}{\beta_1} \ln\hat a_Q \,&\equiv&\,
\beta_1 \Big( \ln\frac{Q}{\Lambda} + \frac{C}{2} \Big) \nn \\
&& \hspace{-18mm} \,=\, \frac{1}{a_Q} + \frac{\beta_1}{2}\,C +
\frac{\beta_2}{\beta_1}\ln a_Q - \beta_1 \!\int\limits_0^{a_Q}\,
\frac{{\rm d}a}{\tilde\beta(a)} \,.
\end{eqnarray}
In perturbation theory, Eq.~\eqn{ahat} should be interpreted in an iterative
sense. Evidently, $\ah_Q$ is a function of $C$ but, for notational simplicity,
we will not make this dependence explicit. One should remark that a combination
similar to~\eqn{ahat}, but without the logarithmic term on the left-hand side,
was already discussed in Refs.~\cite{byz92,ben93}. However, without this term, 
an unwelcome logarithm of $a_Q$ remains in the perturbative relation between
the couplings $\ah_Q$ and $a_Q$. This non-analytic term is avoided by the
construction of Eq.~\eqn{ahat}. 

\begin{figure}
\includegraphics[width=0.47\textwidth]{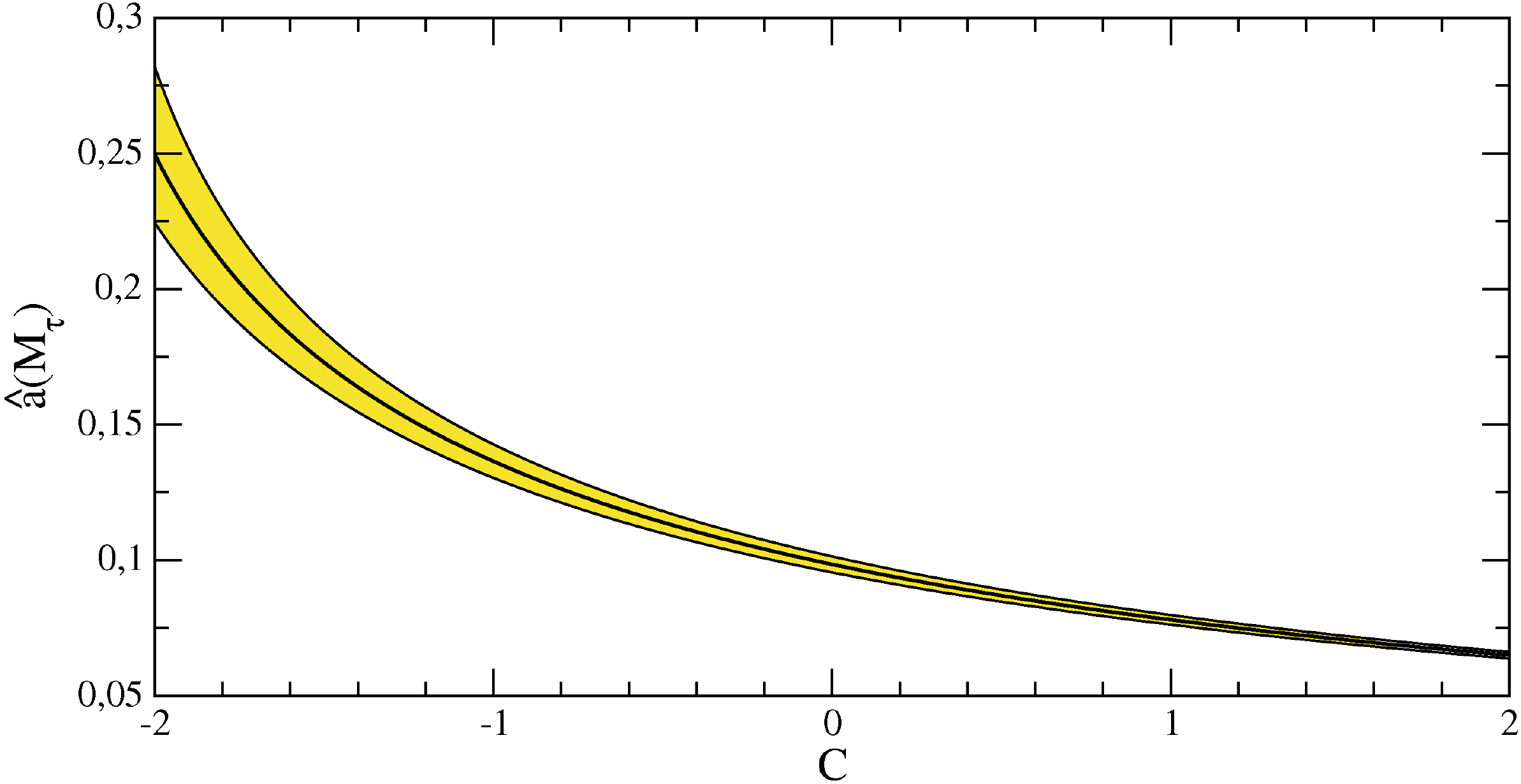}
\caption{The coupling $\ah(M_\tau)$ according to Eq.~\eqn{ahat} as a function
of $C$, and for the $\MSb$ input value $\alpha_s(M_\tau)=0.316(10)$. The
yellow band corresponds to the $\alpha_s$ uncertainty.\label{fig1}}
\end{figure}

In Fig.~\ref{fig1}, we display the coupling $\ah$ according to Eq.~\eqn{ahat}
as a function of $C$. Since in this letter we focus on hadronic $\tau$ decays,
as our initial $\MSb$ input we employ $\alpha_s(M_\tau)=0.316(10)$, which
results from the current PDG average $\alpha_s(M_Z)=0.1181(13)$ \cite{pdg14}.
The yellow band corresponds to the variation within the $\alpha_s$
uncertainties.  Below roughly $C=-2$, the relation between $\ah$ and the
$\MSb$ coupling ceases to be perturbative and breaks down.

The perturbative relations between the coupling $\ah$ and $a$ in a particular
scheme can straightforwardly be deduced from Eq.~\eqn{ahat}. Taking $a$ as well
as the corresponding $\beta$-function coefficients in the $\MSb$ scheme, and
for three quark flavors, $N_f=3$, the expansions read, 
\begin{eqnarray}
\label{ahatofa}
\ah(a) \,&=&\, a - \sfrac{9}{4}\,C\,a^2 - \big( \sfrac{3397}{2592} + 4 C -
\sfrac{81}{16}\,C^2 \big) a^3 - \big( \sfrac{741103}{186624} \nn \\
&& \hspace{-9mm} +\, \sfrac{233}{192}\,C - \sfrac{45}{2}\,C^2 +
\sfrac{729}{64}\,C^3 + \sfrac{445}{144}\zeta_3 \big) a^4 -
\big(\sfrac{727240925}{80621568}  \nn \\ 
\svs
&& \hspace{-9mm}-\,\sfrac{869039}{41472}\,C - \sfrac{26673}{512}\,C^2 +
\sfrac{351}{4}\,C^3 - \sfrac{6561}{256}\,C^4 - \sfrac{445}{32}\zeta_3 C \nn \\ 
\svs
&& \hspace{-9mm} +\,\sfrac{10375693}{373248}\zeta_3 - \sfrac{1335}{256}\zeta_4
- \sfrac{534385}{20736}\zeta_5 \big) a^5 + \cO(a^6) \,,
\end{eqnarray}
and 
\begin{eqnarray}
\label{aofahat}
a(\ah) \,&=&\, \ah + \sfrac{9}{4}\,C\,\ah^2 + \big( \sfrac{3397}{2592} + 4 C +
\sfrac{81}{16}\,C^2 \big) \ah^3 + \big( \sfrac{741103}{186624} \nn \\
\svs
&& \hspace{-9mm} +\, \sfrac{18383}{1152}\,C + \sfrac{45}{2}\,C^2 +
\sfrac{729}{64}\,C^3 + \sfrac{445}{144}\zeta_3 \big) \ah^4 +
\big(\sfrac{1142666849}{80621568} \nn \\ 
\svs
&& \hspace{-9mm} +\,\sfrac{1329359}{20736}\,C + \sfrac{28623 }{256}\,C^2 +
\sfrac{351}{4}\,C^3 + \sfrac{6561}{256}\,C^4 + \sfrac{445}{16}\zeta_3 C \nn \\
\svs 
&& \hspace{-9mm} +\,\sfrac{10375693}{373248}\zeta_3 - \sfrac{1335}{256}\zeta_4
- \sfrac{534385}{20736}\zeta_5 \big) \ah^5 + \cO(\ah^6) \,,
\end{eqnarray}
where $\zeta_i\equiv \zeta(i)$ stands for the Riemann $\zeta$-function.

The running of the coupling $\ah$ can also be deduced from Eq.~\eqn{ahat}. To
this end, one first has to derive its $\beta$-function which is found to have
the simple form
\begin{equation}
\label{betahat}
-\,Q\,\frac{{\rm d}\ah_Q}{{\rm d}Q} \,\equiv\, \hat\beta(\ah_Q) \,=\,
\frac{\beta_1 \ah_Q^2}{\left(1 - \sfrac{\beta_2}{\beta_1}\, \ah_Q\right)} \,.
\end{equation}
As is seen explicitly, it only depends on the scheme-invariant $\beta$-function
coefficients $\beta_1$ and $\beta_2$. It may also be remarked that the only
non-trivial zero of $\hat\beta(\ah)$ arises in the case of $\beta_1=0$.
Integrating the RGE \eqn{betahat} yields
\begin{equation}
\label{ahatrun}
\frac{1}{\ah_Q} \,=\, \frac{1}{\ah_\mu} + \frac{\beta_1}{2}\ln\frac{Q^2}{\mu^2}
- \frac{\beta_2}{\beta_1}\ln\frac{\ah_Q}{\ah_\mu} \,.
\end{equation}
Again, this implicit equation for $\ah_Q$ can either be solved iteratively,
to provide a perturbative expansion, or numerically.

\vspace{2mm}

As our first application of the coupling $\wh\alpha_s$, we investigate the
perturbative series of the Adler function, $D(a_Q)$~\cite{adl74,bck08}. To this
end, it is convenient to define the reduced Adler function $\wh D(a_Q)$ as
\begin{eqnarray}
\label{Dhat}
4\pi^2 D(a_Q) - 1 \,&\equiv&\, \wh D(a_Q) \,=\,
\sum_{n=1}^\infty c_{n,1} a_Q^n \nn \\
&& \hspace{-24mm} \,=\, a_Q + 1.640\,a_Q^2 + 6.371\,a_Q^3 + 49.08\,a_Q^4 +\ldots
\end{eqnarray}
We adopt the notation of Ref.~\cite{bj08}, with numerical coefficients in the
$\MSb$ scheme and for $N_f=3$. The renormalization scale logarithms $\ln(Q/\mu)$
appearing in the Adler function have been resummed with the choice $\mu=Q$.

\begin{figure}
\includegraphics[width=0.46\textwidth]{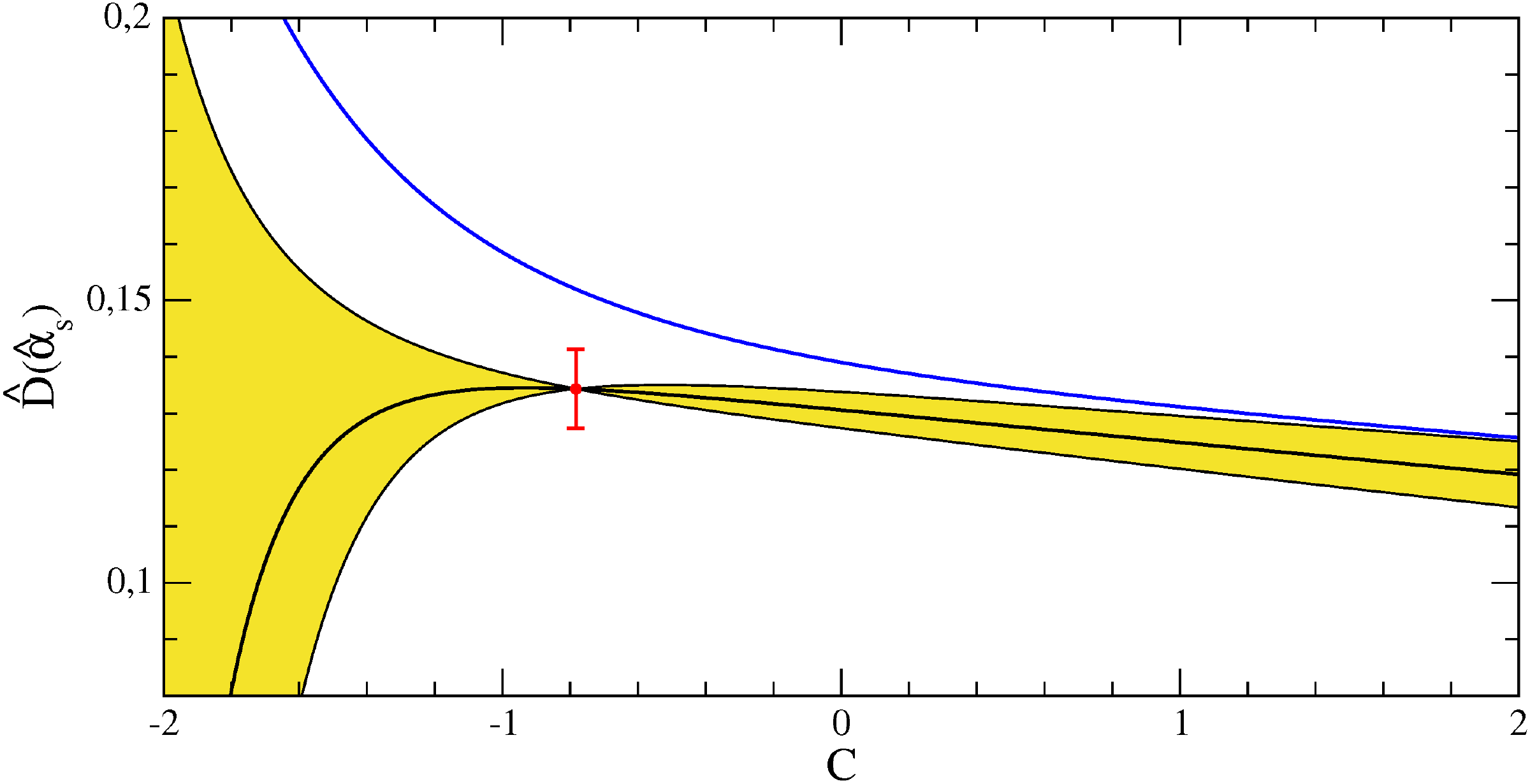}
\caption{$\wh D(\ah_{M_\tau})$ of Eq.~\eqn{Dhatah} as a function of $C$. The
yellow band arises from either removing or doubling the fifth-order term. In
the red dot, the $\cO(\ah^5)$ vanishes, and $\cO(\ah^4)$ is taken as the
uncertainty. For further explanation, see the text.\label{fig2}}
\end{figure}

Using the relation \eqn{aofahat}, we rewrite the expansion \eqn{Dhat} for
$\wh D$ in terms of the $C$-scheme coupling $\ah_Q$, resulting in
\begin{eqnarray}
\label{Dhatah}
\wh D(\ah_Q) \,&=&\, \ah_Q + (1.640 + 2.25 C)\,\ah_Q^2 \nn \\
\svs
&& \hspace{-1cm} +\, (7.682 + 11.38 C + 5.063 C^2)\,\ah_Q^3 \\
\svs
&& \hspace{-1cm} +\, (61.06 + 72.08 C + 47.40 C^2 + 11.39 C^3)\,\ah_Q^4
+ \ldots \nn
\end{eqnarray}
A graphical representation of Eq.~\eqn{Dhatah} is provided in Fig.~\ref{fig2},
where $\wh D(\ah_{M_\tau})$ is plotted as a function of $C$. The yellow band
this time corresponds to an error estimate from the fifth-order contribution.
The required coefficient has been taken to be $c_{5,1}=283$, as estimated in
Ref.~\cite{bj08}. The yellow band then arises by either removing or doubling
the $\cO(\ah^5)$ term. Generally, it is observed that around $C\!\approx\!-1$,
a region of stability with respect to the $C$-variation emerges. For comparison,
the blue line corresponds to using $c_{5,1}=566$ and still doubling the
$\cO(\ah^5)$ correction. Then, no region of stability is found which seems to
indicate that such large values of $c_{5,1}$ are disfavored. In the red dot,
where $C=-0.783$, the $\cO(\ah^5)$ vanishes, and the $\cO(\ah^4)$ correction,
which is the last included non-vanishing term, has been employed as a
conservative  uncertainty, in the spirit of asymptotic expansions. Numerically,
we find
\begin{equation}
\label{Dhoa5zero}
\wh D(\ah_{M_\tau},C=-0.783) \,=\, 0.1343 \pm 0.0070 \pm 0.0067 \,,
\end{equation}
where the second error originates from the uncertainty in $\alpha_s(M_\tau)$.
The result \eqn{Dhoa5zero} may be compared to the direct $\MSb$ prediction
\eqn{Dhat}, which reads
\begin{equation}
\label{DhatMSb}
\wh D(a_{M_\tau}) \,=\, 0.1316 \pm 0.0029 \pm 0.0060 \,.
\end{equation}
Here, the first error is obtained by removing or doubling $c_{5,1}$, and the
second error again corresponds to the $\alpha_s$ uncertainty.

A final comparison of \eqn{Dhoa5zero} and \eqn{DhatMSb} may be performed with
the Adler function model that was put forward in Ref.~\cite{bj08}, and which is
based on general knowledge of the renormalon structure for the Borel transform
of $\wh D(a_Q)$. Within this model, one obtains
\begin{equation}
\label{DhatBM}
\wh D(a_{M_\tau}) \,=\, 0.1354 \pm 0.0127 \pm 0.0058 \,.
\end{equation}
In this case, the first uncertainty results from estimates of the perturbative
ambiguity that arises from the renormalon singularities. It is seen that this
uncertainty is much bigger than the one of \eqn{DhatMSb} and still larger
than the one of \eqn{Dhoa5zero}. Therefore, we conclude that the higher-order
uncertainty of \eqn{DhatMSb} appears to be underestimated, while
Eq.~\eqn{Dhoa5zero} seems to provide a more realistic account of the resummed
series. Interestingly enough, also its central value is closer to the Borel
model result.

\vspace{2mm}

Now, we turn to the perturbative expansion for the total $\tau$ hadronic width.
The central observable is the ratio $R_\tau$ of the total hadronic branching
fraction to the electron branching fraction. It can be parameterized as
\begin{equation}
R_\tau \,=\, 3\, S_{\rm EW} (|V_{ud}|^2 + |V_{us}|^2)\, ( 1 + \delta^{(0)}
+ \cdots) \,,
\end{equation}
where $S_{\rm EW}$ is an electroweak correction and $V_{ud}$ as well as $V_{us}$
CKM matrix elements. Perturbative QCD is encoded in $\delta^{(0)}$ (see
Refs.~\cite{bnp92,bj08} for  details) and the ellipsis indicate further small
subleading corrections. For $\delta^{(0)}$ a complication arises, because it is
calculated from a contour integral in the complex energy plane. On the other
hand, we seek to resum the scale logarithms $\ln(Q/\mu)$, and the perturbative
prediction depends on whether those logs are resummed before or after performing
the contour integration. The first choice is called contour-improved
perturbation theory (CIPT) \cite{dp92} and the second fixed-order perturbation
theory (FOPT).

In FOPT, the perturbative series of $\delta^{(0)}(a_Q)$ in terms of the $\MSb$
coupling $a_Q$ is given by \cite{bck08,bj08}
\begin{equation}
\label{del0}
\delta_{\rm FO}^{(0)}(a_Q) \,=\,
a_Q + 5.202\,a_Q^2 + 26.37\,a_Q^3 + 127.1\,a_Q^4 +\ldots
\end{equation}
On the other hand, in the $C$-scheme coupling $\ah_Q$, the expansion for
$\delta^{(0)}(\ah_Q)$ reads
\begin{eqnarray}
\label{del0ah}
\delta_{\rm FO}^{(0)}(\ah_Q) \,&=&\, \ah_Q + (5.202 + 2.25 C)\,\ah_Q^2 \nn \\
\svs
&& \hspace{-1cm} +\, (27.68 + 27.41 C + 5.063 C^2)\,\ah_Q^3 \\
\svs
&& \hspace{-1cm} +\, (148.4 + 235.5 C + 101.5 C^2 + 11.39 C^3)\,\ah_Q^4
+ \ldots \nn
\end{eqnarray}

\begin{figure}
\includegraphics[height=4.2cm]{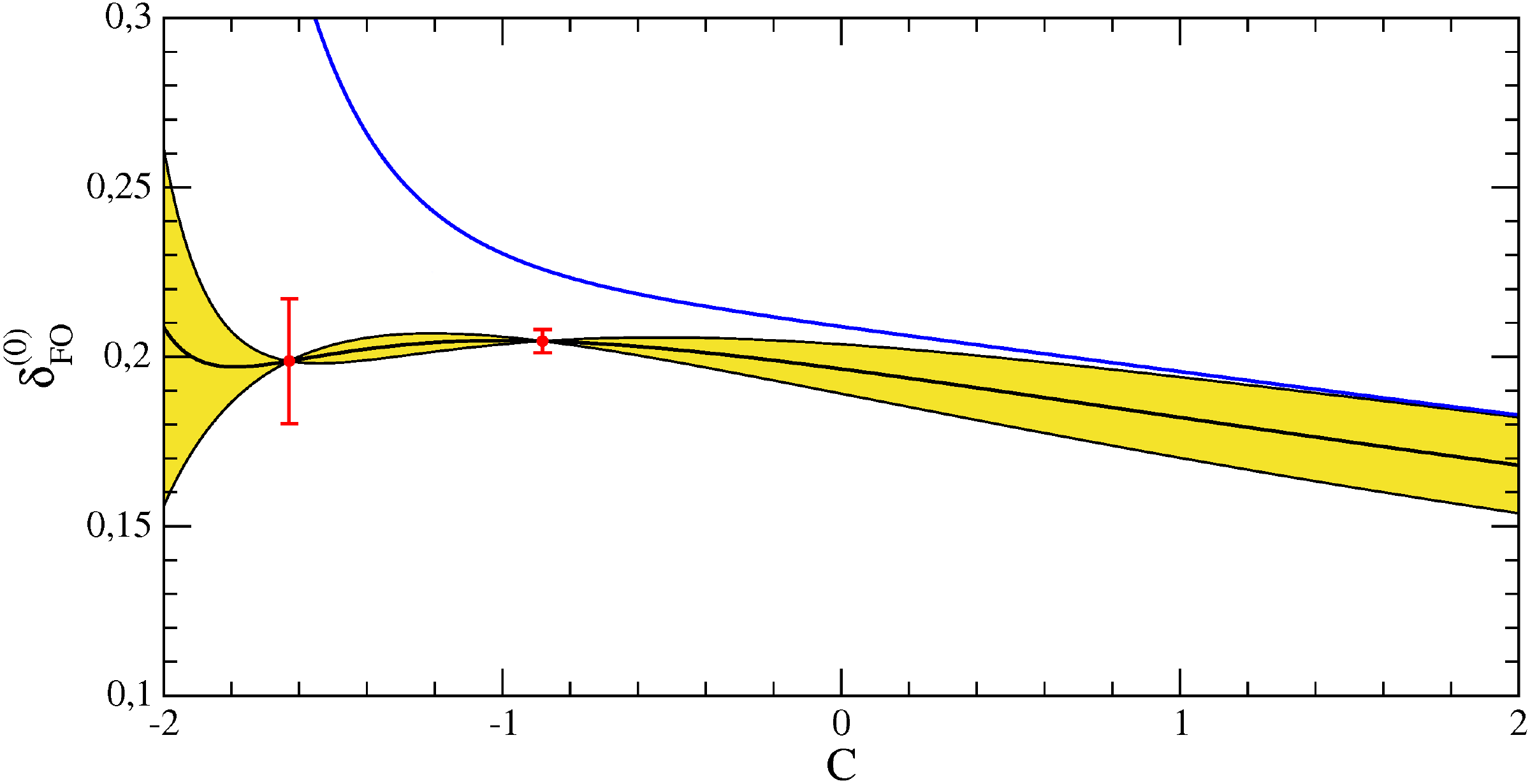}
\caption{$\delta_{\rm FO}^{(0)}(\ah_Q)$ of Eq.~\eqn{del0ah} as a function of
$C$. The yellow band arises from either removing or doubling the fifth-order
term. In the red dots, the $\cO(\ah^5)$ vanishes, and $\cO(\ah^4)$ is taken as
the uncertainty. For further explanation, see the text.\label{fig3}}
\end{figure}

In Fig.~\ref{fig3}, we display $\delta_{\rm FO}^{(0)}(\ah_Q)$ as a function
of $C$. Assuming $c_{5,1}=283$, the yellow band again corresponds to removing
or doubling the $\cO(\ah^5)$ term. Like for $\wh D(\ah)$, a nice plateau is
found for $C\approx -1$. Taking $c_{5,1}=566$ and then doubling the $\cO(\ah^5)$
results in the blue curve that does not show stability. Hence, this scenario
again is disfavored. In the red dots, which lie at $C=-0.882$ and $C=-1.629$,
the $\cO(\ah^5)$ correction vanishes, and the $\cO(\ah^4)$ term is taken as
the uncertainty. The point to the right has a substantially smaller error, and
yields
\begin{equation}
\label{del0oa5zero}
\delta_{\rm FO}^{(0)}(\ah_{M_\tau},C=-0.882) \,=\,
0.2047 \pm 0.0034 \pm 0.0133 \,.
\end{equation}
Once more, the second error covers the uncertainty of $\alpha_s(M_\tau)$.
In this case, the direct $\MSb$ prediction of Eq.~\eqn{del0} is found to be
\begin{equation}
\label{del0MSb}
\delta_{\rm FO}^{(0)}(a_{M_\tau}) \,=\, 0.1991 \pm 0.0061 \pm 0.0119 \,.
\end{equation}
This value is somewhat lower, but within $1\,\sigma$ of the higher-order
uncertainty. Comparing, on the other hand, to the Borel model (BM) result
of \cite{bj08}, which is given by
\begin{equation}
\label{del0BM}
\delta_{\rm BM}^{(0)}(a_{M_\tau}) \,=\, 0.2047 \pm 0.0029 \pm 0.0130 \,,
\end{equation}
it is found that \eqn{del0oa5zero} and \eqn{del0BM} are surprisingly similar.
In both cases, the parametric $\alpha_s$ uncertainty is substantially larger
than the higher-order one  -- especially given the recent increase in the
$\alpha_s$ uncertainty provided by the PDG \cite{pdg14} -- which underlines
the good potential of $\alpha_s$ extractions from hadronic $\tau$ decays.

\begin{figure}
\includegraphics[height=4.2cm]{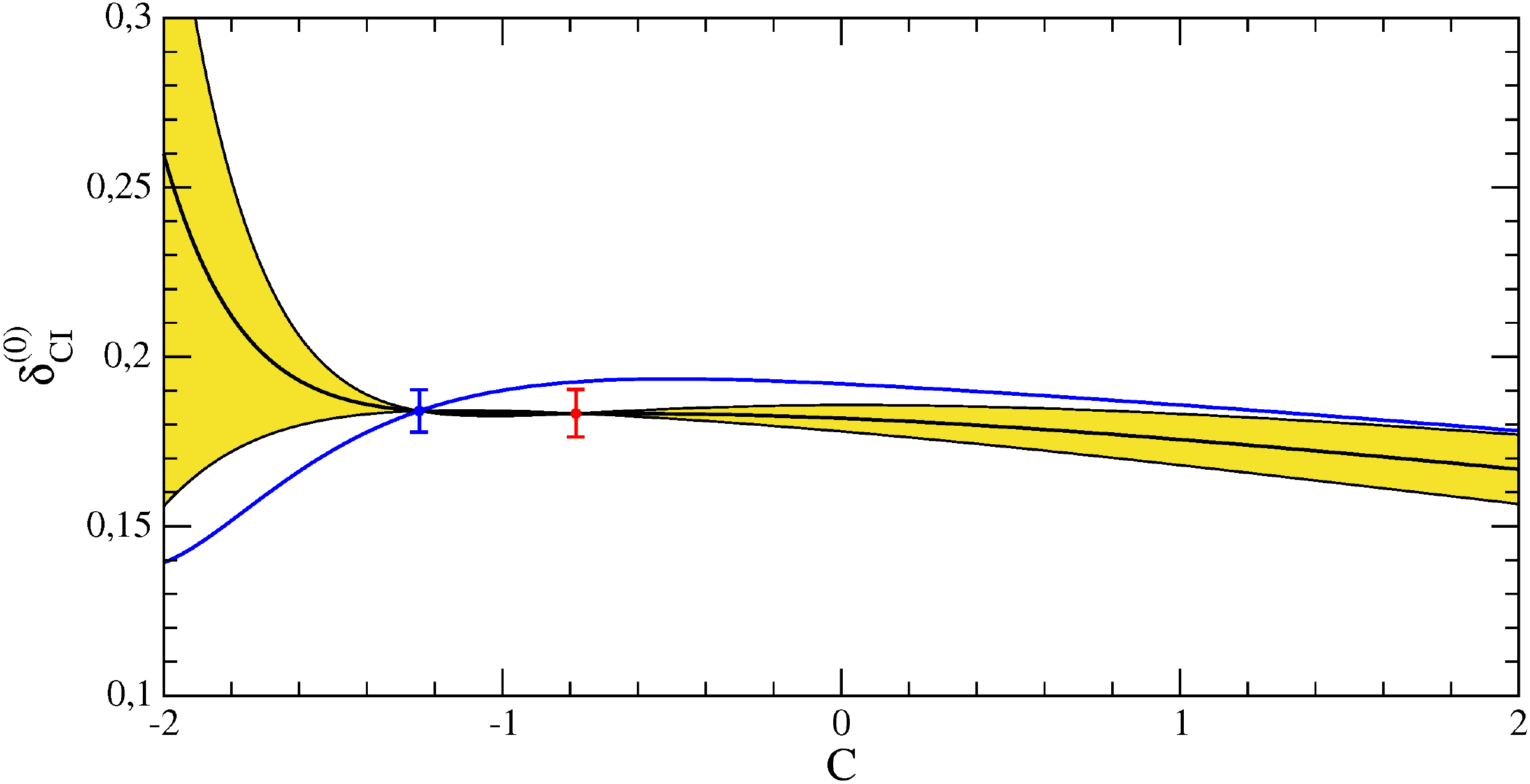}
\caption{$\delta_{\rm CI}^{(0)}(\ah_Q)$ as a function of $C$. The yellow band
arises from either removing or doubling the fifth-order term. In the red and
blue dots, the $\cO(\ah^5)$ vanishes, and $\cO(\ah^4)$ is taken as the
uncertainty. For further explanation, see the text.\label{fig4}}
\end{figure}

In CIPT, contour integrals over the running coupling, Eq.~(\ref{ahatrun}),
have to be computed, and hence the result cannot be given in analytical form.
Graphically, $\delta_{\rm CI}^{(0)}(a_{M_\tau})$ as a function of $C$ is
displayed in Fig.~\ref{fig4}. The general behavior is very similar to FOPT,
with the exception that now also for $c_{5,1}=566$ a zero of the $\cO(\ah^5)$
term is found. This time, both zeros  have similar uncertainties, and employing
the point with smaller error (in blue) yields
\begin{equation}
\label{del0CIoa5z}
\delta_{\rm CI}^{(0)}(\ah_{M_\tau},C=-1.246) \,=\,
0.1840 \pm 0.0062 \pm 0.0084 \,.
\end{equation}
As has been discussed many times in the past (see e.g.~\cite{bj08}) the CIPT
prediction lies substantially below the FOPT results, especially the $C$-scheme
ones, and the Borel model. On the other hand, the parametric $\alpha_s$
uncertainty in CIPT turns out to be smaller.

\vspace{2mm}

In this work, in Eq.~\eqn{ahat}, we have defined a class of QCD couplings
$\ah_Q$, such that the scale running is explicitly scheme invariant, and scheme
changes are parameterized by a single constant $C$. For this reason, we have
termed $\ah_Q$ the $C$-scheme coupling. Scheme transformations correspond to
changes in the QCD scale $\Lambda$.

We have applied the coupling $\ah_Q$ to investigations of the perturbative
series of the reduced Adler function $\wh D$. Our central result is given in
Eq.~\eqn{Dhoa5zero}. Its higher-order uncertainty turned out larger than the
corresponding $\MSb$ prediction \eqn{DhatMSb}, but we consider \eqn{Dhoa5zero}
to be more realistic and conservative.

We also studied the perturbative expansion of the $\tau$ hadronic width,
employing the coupling $\ah_Q$. In this case our central prediction in FOPT
is given in Eq.~\eqn{del0oa5zero}. Surprisingly, the result \eqn{del0oa5zero}
is found very close to the prediction \eqn{del0BM} of the central Borel model
developed in Ref.~\cite{bj08}, hence providing some support for this approach.

The disparity between FOPT and CIPT predictions for $\delta^{(0)}$ is not
resolved by the $C$-scheme.  As is seen from Eq.~\eqn{del0CIoa5z}, the CIPT
result turns out substantially lower (as is the case for the $\MSb$ prediction).
This suggests to return to investigations of Borel models, this time in the
coupling $\ah$, in order to assess the scheme dependence of such models. This
could result in an improved extraction of $\alpha_s$ from hadronic decays of
the $\tau$ lepton.

\vspace{2mm}
\begin{acknowledgments}
Helpful discussions with Martin~Beneke are gratefully acknowledged.
The work of MJ and RM has been supported in part by MINECO Grant number
CICYT-FEDER-FPA2014-55613-P, by the Severo Ochoa excellence program of MINECO,
Grant SO-2012-0234, and Secretaria d'Universitats i Recerca del Departament
d'Economia i Coneixement de la Generalitat de Catalunya under Grant 2014 SGR
1450. DB's is supported by the S\~ao Paulo Research Foundation (FAPESP) grant
15/20689-9, and by CNPq grant 305431/2015-3. 
\end{acknowledgments}

%

\vfill

\end{document}